# Detection of Nonlinear Behavior in Voltage Source Converter Control in Wind Farms Based on Higher-Order Spectral Analysis

Zetian Zheng, *Student Member, IEEE,* and Chen Shen, *Senior Member, IEEE*

*Abstract*—In recent years, the sub-synchronous oscillation (SSO) accidents caused by wind power have received extensive attention. A method is needed to distinguish if nonlinear behavior exists in the recorded equal-amplitude accident waveforms, so that different methods can be adopted to analyze the mechanism of the oscillation. The theory of higher-order statistics (HOS) has become a powerful tool for detection of nonlinear behavior (DNB) in production quality control since 1960s. However, HOS analysis has been applied in mechanical condition monitoring and fault diagnosis, even after being introduced into the power system and wind farms. This paper focuses on the voltage source converter (VSC) control systems in wind farms and tries to detect the nonlinear behavior caused by the bilateral or unilateral saturation hard limits based on HOS analysis. First, the traditional describing function is extended to obtain more frequency domain information, and hereby the harmonic characteristics of bilateral and the unilateral saturation hard limit are studied. Then the bispectrum and trispectrum are introduced as HOS, which are extended into bicoherence and tricoherence spectrums to eliminate the effects from linear parts in the VSC control system. The effectiveness of DNB and classification based on HOS is strictly proved and its detailed calculation and estimation process is illustrated. Finally, the proposed method is demonstrated and further discussed through simulation results.

## I. Introduction

WITH the implementation of the energy development strategy in China, the installation capacity of wind power increases in high speed. However, the stability problems brought about by wind integration are also becoming more and more serious. Among them, the sub-synchronous oscillation (SSO) is the most prominent. After the SSO accidents, the collected accident waveform record tends to be of equal-amplitude because the transition process is very fast. Among current literatures, most scholars analyze the SSO based on the weakly-damped assumption, therefore they linearized the power system dynamic equations and conducted their studies using analysis methods for linear systems. These methods mainly includes the eigenvalue method [1], the dynamic equivalent method [2], [3], the impedance method [4], [5], and the complex torque coefficient method [6], [7] and so on. However, there are nonlinear parts such as hard limits in the control system of wind power, and the induced self-sustained oscillation is also of constant amplitude. So, it is unreasonable to directly analyze the amplitude and frequency of SSO with a linearized method, and adopt corresponding measures to suppress the oscillation. It is necessary to distinguish the oscillation type, linear or nonlinear, from the waveform records before choosing analyzing methods.

Since its emergence in the early 1960s, the theory of higher-order statistics (HOS) has become a powerful analysis tool in condition monitoring and fault diagnosis of mechanical equipment. Its applications mainly include three aspects at the very beginning: harmonic retrieval [8], system identification [9], and feature extraction [10]. Later, researchers applied HOS to detection of nonlinear behavior (DNB) [11]. On the basis of bispectrum, [12] and [13] proposed the definitions of bicoherence spectrum and inverted bispectrum, respectively. Since then, scholars have proposed different statistical indicators based on HOS for DNB.

In the field of power systems, inspired by the above ideas, as early as 1995, researchers have proved that bispectral analysis can be introduced into fault identification and condition monitoring of three-phase induction motors to analyze and identify motor asymmetric faults and stator winding failure [14]. Since then, the research on fault diagnosis using HOS has achieved fruitful results. Reference [15] detected and identified asymmetric faults in induction motors by measuring vibration data and analyzing motor nonlinearity using the bicoherence spectrum. In [16], considering the Gaussian noise and non-Gaussian noise of mechanical signals, a new rolling bearing detection method was proposed, which integrated bispectral analysis and improved ensemble empirical mode decomposition.

In wind farms, [17] used a modulated signal bispectrum detector to diagnose bearing faults of wind turbines for doubly-fed induction generator wind turbines. [18] used bispectral analysis to identify single-point defects in rolling bearings. [19] proposed an improved signal separation method based on the Vold Kalman filter and the HOS analysis for rotating mechanical systems under strong background noise.

This work was supported by the National Natural Science Foundation of China under Grant Nos. U1766206, 51677098, and 51621065. (Corresponding author: Chen Shen.)
Z. Zheng and C. Shen are with the State Key Laboratory of Power Systems, Department of Electrical Engineering, Tsinghua University, Beijing 100084, China (e-mail: zzt_thu@qq.com; shenchen@mail.tsinghua.edu.cn).

However, the current application of bispectral analysis mainly focuses on the mechanical defects of the power system, and there are few studies analyze the nonlinearity of the control system in the power system or specifically in wind farms. Meanwhile, the voltage source converter (VSC) is an essential component of wind turbines and static var generators (SVGs). To apply HOS in the control system of VSCs in wind farms for DNB, four aspects need to be considered:

1) whether the nonlinearity of the VSC control system can still be characterized by HOS and which HOS should be used;

2) how to figure out the characteristics with only the waveform collected at the terminal of the VSC;

3) how to detect the nonlinear behavior caused by the bilateral or unilateral saturation hard limits, which is considered the source of nonlinearity in this paper;

4) how to improve the effectiveness and quality of the spectrum.

This paper attempts to give an analytical proof of the effectiveness of HOS applied in DNB of the VSC control.

The remainder of this paper is organized as follows. Section II extends the traditional describing function and analyzes two types of hard limits. In Section III, HOS is introduced. In Section IV, the VSC control system is modeled and DNB based on HOS is studied and strictly proved. The detailed calculation process for DNB in the VSC control system is described in Section V, and its effectiveness is proved through case studies in Section VI. Finally, conclusions derived from this paper is presented in Section VII.

## II. ANALYSIS OF HARMONIC CHARACTERISTICS OF HARD LIMIT

The describing function (DF) [20] has been effectively used to analyze the characteristics of sustained oscillations (or limit cycles) caused by nonlinearities. In the traditional DF, only the first-order Fourier series of the oscillation is reserved. In DNB, however, the higher-order characteristics are needed to tell apart the sustained oscillations induced by hard limits and negatively damped oscillations induced by incorrectly configured control parameters. So, the DF method is introduced and extended in this section, and the harmonic characteristics of the bilateral and unilateral saturation hard limit are discussed.

### A. The Extended DF

As shown in Fig. 1, assume that the input signal of the studied nonlinear part is a sinusoidal signal which is described as

$$x(t) = A\sin\omega t \quad (1)$$

where $A$ and $\omega$ are the amplitude and the frequency of the input sinusoidal signal, respectively.

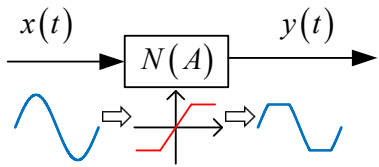

Fig. 1. A typical hard limit with a sinusoidal input

The output $y(t)$ of the hard limit is a periodic non-sinusoidal signal, which can be expanded into a Fourier series as

$$y(t) = A_0 + \sum_{n=1}^{\infty}(A_n \cos n\omega t + B_n \sin n\omega t) \quad (2)$$

where $A_0$ is the magnitude of the DC component; $A_n$ and $B_n$ are the cosine part and sine part of the magnitude of the n-th Fourier harmonics, respectively.

In the traditional DF-based analysis method, the nonlinear part is considered oddly symmetrical and the linear part of the system is considered to be low-pass. Then $A_0 = 0$ is derived and $y(t)$ can be approximated as

$$y(t) \approx y_1(t) = A_1 \cos \omega t + B_1 \sin \omega t = Y_1 e^{j\varphi_1 t} \quad (3)$$

where

$$A_1 = \frac{1}{\pi}\int_0^{2\pi} y(t)\cos\omega t\, d(\omega t) \quad Y_1 = \sqrt{A_1^2 + B_1^2}$$
$$B_1 = \frac{1}{\pi}\int_0^{2\pi} y(t)\sin\omega t\, d(\omega t) \quad \varphi_1 = \arctan\frac{B_1}{A_1} \quad (4)$$

The ratio of the first-order Fourier series of $y(t)$ and the magnitude of the input signal is defined as the DF of the nonlinear part:

$$N(A) = \frac{Y_1}{A}e^{j\varphi_1} \quad (5)$$

To extend the DF method, the higher-order harmonics $A_n$ and $B_n$ in (2) are calculated as

$$A_n = \frac{1}{\pi}\int_0^{2\pi} x(t)\cos n\omega t\, d(\omega t)$$
$$B_n = \frac{1}{\pi}\int_0^{2\pi} x(t)\sin n\omega t\, d(\omega t) \quad (6)$$

### B. Bilateral Saturation Hard Limit

As shown in Fig. 2(a), when a sine wave goes through a bilateral saturation hard limit in time domain, its upper and lower part exceeding the limit is set at the limit value, which can be described as

$$x(t) = A\sin\omega t$$
$$y(t) = \begin{cases} -a & x(t) < -a \\ x(t) & -a \leq x(t) \leq a \\ a & x(t) > a \end{cases} \quad (7)$$

where $A$ and $\omega$ are the amplitude and the frequency of the input sinusoidal signal, and $a\ (>0)$ is the upper limit of the hard limit.

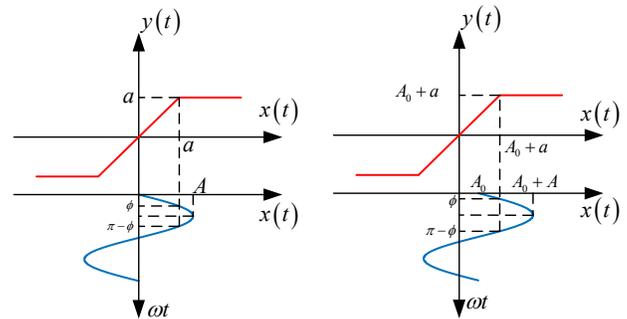

(a) bilateral-saturation  (b) unilateral-saturation

Fig. 2. Time-domain characteristics of bilateral-saturation and unilateral-saturation hard limits

The bilateral hard limit is oddly symmetrical, and the output periodic signal is an odd function, so the coefficients of the DC component and the cosine components in the Fourier series are 0, i.e., $A_n = 0 \ (n = 0,1,2,...)$. According to (4), the sine fundamental component of the output can be derived as

$$B_1 = \frac{1}{\pi}\int_0^{2\pi} x(t)\sin\omega t \, d(\omega t) \quad (8)$$
$$= \frac{2}{\pi}\left(\int_0^{\phi} A\sin^2\omega t \, d(\omega t) + \int_{\phi}^{\pi-\phi} a\sin\omega t \, d(\omega t) + \int_{\pi-\phi}^{\pi} A\sin^2\omega t \, d(\omega t)\right)$$
$$= \frac{2A}{\pi}\left(\arcsin\frac{a}{A} + \frac{a}{A}\sqrt{1-\left(\frac{a}{A}\right)^2}\right)$$

The n-th components are calculated similarly and the results show that only

$$B_{2n+1} \neq 0 \quad (n = 0,1,2,\cdots) \quad (9)$$

The detailed calculation results can be referred to Table A in the appendix.

### C. Unilateral Saturation Hard Limit

As shown in Fig. 2 (b), when a sine wave goes through a unilateral saturation hard limit in the time domain, its upper (or lower, depending on the actual situation) part exceeding the limit is set at the limit value, which can be described as

$$x(t) = A\sin\omega t$$
$$y(t) = \begin{cases} x(t) & x(t) \leq A_0 + a \\ A_0 + a & x(t) > A_0 + a \end{cases} \quad (10)$$

where $A$, $A_0$ and $\omega$ are the amplitude, offset and the frequency of the input sinusoidal signal, and $A_0 + a$ is the upper limit of the hard limit.

Similar to II.B, the n-th components can be calculated and the results show that

$$\begin{aligned} A_{2n} &\neq 0 \quad (n=0,1,2,\cdots) \\ B_{2n+1} &\neq 0 \quad (n=0,1,2,\cdots) \end{aligned} \quad (11)$$

The detailed calculation results can be referred to Table B in the appendix.

## III. HOS ANALYSIS

In order to analyze the harmonic characteristics of nonlinear parts, the definition of HOS is introduced in this section. The eigenfunction method is one of the important tools of statistical analysis, which can easily lead to the definition of higher-order moments and higher-order cumulants.

The first joint eigenfunction of $k$ continuous random variables $x_1, \cdots, x_k$ is defined as

$$\Phi(\omega_1,\cdots,\omega_k) \doteq E\left\{e^{j(\omega_1 x_1+\cdots+\omega_k x_k)}\right\} \quad (12)$$
$$= \int_{-\infty}^{\infty}\cdots\int_{-\infty}^{\infty} f(x_1,\cdots,x_k)e^{j(\omega_1 x_1+\cdots+\omega_k x_k)}dx_1\cdots dx_k$$

where $f(\cdot)$ is the probability density function.

The k-order moments and k-order cumulants of $k$ random variables are derived respectively as

$$E(x_1,\cdots,x_k) = (-j)^k \left.\frac{\partial^k \Phi(\omega_1,\cdots,\omega_k)}{\partial\omega_1\cdots\partial\omega_k}\right|_{\omega_1=\cdots=\omega_k=0} \quad (13)$$

$$cum(x_1,\cdots,x_k) = (-j)^k \left.\frac{\partial^k \ln\Phi(\omega_1,\cdots,\omega_k)}{\partial\omega_1\cdots\partial\omega_k}\right|_{\omega_1=\cdots=\omega_k=0} \quad (14)$$

For a stationary continuous random signal $x(t)$, set $x_1 = x(t), x_2 = x(t+\tau_1),\cdots, x_k = x(t+\tau_{k-1})$ in (14), then the kth-order cumulant of the random signal $x(t)$ is represented as

$$c_{kx}(\tau_1,\cdots,\tau_{k-1}) = cum[x(t),x(t+\tau_1),\cdots,x(t+\tau_{k-1})] \quad (15)$$

The k-order cumulant spectrum is defined as the (k-1)-dimensional discrete Fourier transform of the k-order cumulant, which is calculated as

$$S_{kx}(\omega_1,\cdots,\omega_{k-1}) = \sum_{\tau_1=-\infty}^{\infty}\cdots\sum_{\tau_{k-1}=-\infty}^{\infty} c_{kx}(\tau_1,\cdots,\tau_{k-1})e^{-j(\omega_1\tau_1+\cdots+\omega_{k-1}\tau_{k-1})} \quad (16)$$

Generally, the higher-order cumulant spectrum is simply referred to as the higher-order spectrum. In particular, the third-order spectrum $S_{3x}(\omega_1,\omega_2)$ is called bispectrum because it is an energy spectrum of two frequencies, and is represented by $B(\omega_1,\omega_2)$ in this paper. Likewise, the fourth-order spectrum $S_{4x}(\omega_1,\omega_2,\omega_3)$ is referred to as the trispectrum, herein denoted by $T(\omega_1,\omega_2,\omega_3)$.

## IV. DNB OF VSC CONTROL SYSTEM

### A. Modeling of VSC Control System

A typical direct-drive wind farm has 30–60 generators. Several direct-drive permanent magnet synchronous generators (PMSGs) are connected, forming a string structure. Then several strings are attached to PCC and finally to the main grid through a series of boosting transformers and transmission lines (which is equivalent to a set of impedances) [21]. Fig. 3 is the schematic of a typical direct-drive wind generator, which consists of a wind turbine, a PMSG, and a full-power converter (including machine- and grid-side converters).

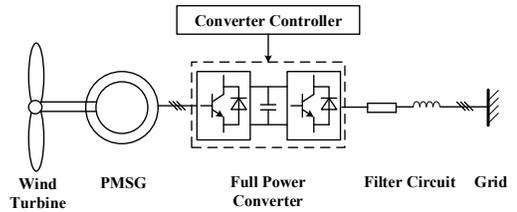

Fig. 3. Structure of a typical direct-drive wind generator.

As the machine-side converter adopts maximum power tracking control, its interaction with the grid is small. As noted by [22] and [23], the grid-related oscillation dynamics strongly depend on the DC capacitor and grid-side converter but are weakly affected by the wind turbine, PMSG, and machine-side converter. Therefore, the machine-side component (including the wind turbine, PMSG, and machine-side converter) is equivalent to a power source that outputs wind power received by the wind turbine and the grid-side converter is modeled as a VSC with its corresponding control system.

Meanwhile, the static reactive power compensation equipment such as static VAR generator (SVG) is usually installed in wind farms. The SVG model in this paper adopts a double closed-loop control strategy. The d-axis control loop stabilizes the DC bus voltage, and the q-axis control loop varies according to the control mode. When a SVG operates in a constant-voltage control mode, the control target of this loop is the terminal voltage; when it operates in constant reactive power control mode, the control target of this loop is output reactive power. The SVG is hereby modeled as a VSC with its corresponding control system.

So far, we obtain a unified VSC control system for both PMSGs and SVGs in wind farms. The only difference is the choice of the control targets in the d-axis and the q-axis control loop, as is shown in Fig. 4. In this paper, 4 hard limits are considered as the nonlinear parts: 2 in the inner control loop of current and 2 in the outer control loop of voltage.

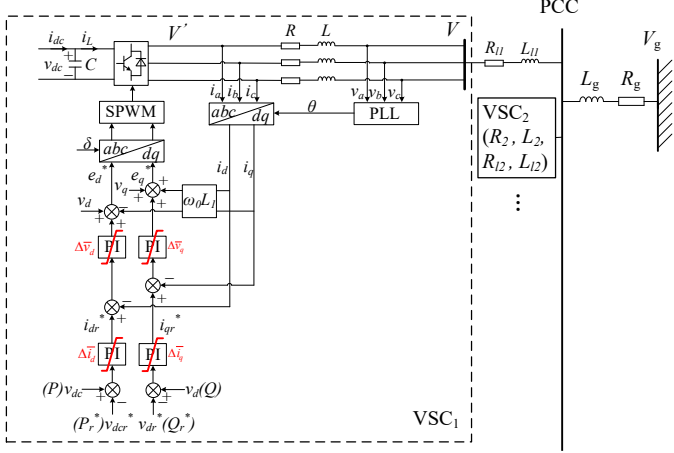

Fig. 4. Structure of VSC control systems.

The meanings of the symbols in Fig. 4 are as follows: $v_{dc}$ (or $P$) and $v_{dcr}^*$ (or $P_r^*$) are the measurement and the reference of the DC bus voltage (or DC power), respectively; $v_d$ (or $Q$) and $v_{dr}^*$ (or $Q_r^*$) are the measurement and the reference of the d-axis terminal voltage (or the output reactive power), respectively; $i_d$ and $i_q$ are the $d$ axis component and $q$ axis component of the output current of the grid-side converter, respectively; $e_d$ and $e_q$ are the $d$ axis component and $q$ axis component of the fundamental output voltage of the grid-side converter, respectively; $V'$, $V$ and $V_g$ are the phase voltage amplitude of the terminal, the PCC and the grid, respectively; $\delta$ is the output angle of the PLL; $R_1$ and $L_1$ (or $R_2$ and $L_2$) are the equivalent resistance and inductance of the transformer and the transmission line between the terminal and PCC, respectively; $R_g$ and $L_g$ are the equivalent resistance and inductance of the transmission line between PCC and the grid, respectively.

B. Elimination of the Effects from Linear Parts

While the nonlinear parts take effect inside the control system, the accident waveform record collected from the phasor measurement unit (PMU) offers only the voltage and current information at the terminal of the VSC. Therefore, it's necessary to derive the relationship between the HOS of the hard limit output and that of the terminal electrical quantities.

When the hard limit in the d-axis inner control loop of current takes effect, the output $\bar{v}_d$ is produced by the nonlinear part after PI. Assume the phase-locked loop (PLL) performs well and the VSC keeps synchronized with the system, then the d-axis frame model of the main circuit is written as

$$\Delta v'_d - \Delta v_d = (sL + R)\Delta i_d - \omega_0 L \Delta i_q \quad (17)$$

Ignoring the dynamics of the PWM, assume $v'_d = e_d^*$, then the relationship between $\bar{v}_d$ and $v'_d$ is derived as

$$\Delta v'_d = \Delta \bar{v}_d - \omega_0 L \Delta i_q + \Delta v_d \quad (18)$$

Combining (17) and (18), we get the relationship between $i_d$ and $\bar{v}_d$ as

$$\Delta i_d = \frac{1}{sL+R}\Delta \bar{v}_d \quad (19)$$

Similarly, when the hard limit in the q-axis inner control loop of current takes effect, the relationship between $i_q$ and $\bar{v}_q$ stands as

$$\Delta i_q = \frac{1}{sL+R}\Delta \bar{v}_q \quad (20)$$

When the hard limit in the d-axis outer control loop of voltage takes effect, the output $\bar{i}_d$ is produced by the nonlinear part after PI. (18) turns into

$$\Delta v'_d = G_i(s)\cdot\left(\Delta \bar{i}_d - \Delta i_d\right) - \omega_0 L \Delta i_q + \Delta v_d \quad (21)$$

Combining (17) and (21), we get the relationship between $i_d$ and $\bar{i}_d$ as

$$\Delta i_d = \frac{G_i}{sL+R+G_i}\Delta \bar{i}_d \quad (22)$$

where $G_i$ is the transfer function of the inner control loop of current. Similarly, when the hard limit in the q-axis outer control loop of voltage takes effect, the relationship between $i_q$ and $\bar{i}_q$ stands as

$$\Delta i_q = \frac{G_i}{sL+R+G_i}\Delta \bar{i}_q \quad (23)$$

Therefore, combining (19), (20), (22) and (23), the output of the nonlinear part in the control system can always be obtained from the current measured at the terminal after going through a linear part. To eliminate the effects of linear parts on the HOS of the terminal current, consider the system in Fig. 5, with the input $e(n)$ and the output $y(n)$. $H(z)$ represents a linear time-invariant part.

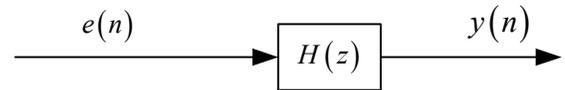

Fig. 5. A typical linear system.

Combining the definition and properties of the HOS, it can be proved that the following relationship exists between the HOS of $e(n)$ and $y(n)$:

$$S_{ky}(\omega_1,\cdots,\omega_{k-1}) = S_{ke}(\omega_1,\cdots,\omega_{k-1})H(\omega_1)\cdots H(\omega_{k-1})H^*(\omega_1+\cdots+\omega_{k-1}) \quad (24)$$

where $S_{ke}(\omega_1,\cdots,\omega_{k-1})$ and $S_{ky}(\omega_1,\cdots,\omega_{k-1})$ are the k-order cumulant spectrum of $e(n)$ the $y(n)$, respectively, and $H(\omega)$ is the continuous transfer function of $H(z)$. Let $k = 2,3$ in (24), we get

$$P_y(\omega) = P_e(\omega)|H(\omega)|^2 \quad (25)$$

$$B_y(\omega_1,\omega_2) = B_e(\omega_1,\omega_2)H(\omega_1)H(\omega_2)H^*(\omega_1+\omega_2) \quad (26)$$

Define $bic(\omega_1,\omega_2)$ as the bicoherence at the frequency pair $(\omega_1,\omega_2)$, which is calculated as

$$bic(\omega_1,\omega_2) = \left|\frac{B(\omega_1,\omega_2)}{\sqrt{P(\omega_1)P(\omega_2)P(\omega_1+\omega_2)}}\right| \quad (27)$$

Combining (25), (26) and (27), it can be proved that

$$bic_y(\omega_1,\omega_2) = bic_e(\omega_1,\omega_2) \quad (28)$$

Define tricoherence $tric(\omega_1, \omega_2, \omega_3)$ as

$$tric(\omega_1,\omega_2,\omega_3) = \frac{|T(\omega_1,\omega_2,\omega_3)|}{\sqrt{P(\omega_1+\omega_2+\omega_3)P(\omega_1)P(\omega_2)P(\omega_3)}} \quad (29)$$

Similarly, it can be proved that $tric_y(\omega_1, \omega_2, \omega_3) = tric_e(\omega_1, \omega_2, \omega_3)$.

From the above deduction, it is clear that the linear part does not change the bicoherence and the tricoherence of the system. Combining (19), (20), (22) and (23), by measuring the waveform of $i_d$ and $i_q$ at the terminal of the VSC and performing HOS analysis on them, the nonlinearity of the VSC system can be detected.

### C. Bicoherence Spectrum and Unilateral Saturation Hard Limit Detection

From the results in II.C, when the self-sustained oscillation occurs, the output $y$ of the unilateral saturation hard limit contains the second harmonic of the oscillation frequency with the same phase as that of the fundamental frequency. Without loss of generality, let its initial phase be 0, i.e.,

$$y(t) = B_1 \sin 2\pi f t + B_2 \sin 2\pi \cdot 2 f t \quad (30)$$

where $f$ is the oscillation frequency. $B_1$ and $B_2$ are the cosine parts of the magnitude of the fundamental and second Fourier harmonics, respectively.

The Fourier transform is performed twice on the second-order autocorrelation function of $y(t)$ and then the bispectrum of $y(t)$ can be derived as

$$B_y(\omega_1,\omega_2) = \int_{-\infty}^{\infty}\int_{-\infty}^{\infty} f\int_0^{1/f} y(t)y(t+\tau_1)y(t+\tau_2)dt \times e^{-j2\pi(\omega_1\tau_1+\omega_2\tau_2)}d\tau_1 d\tau_2 \quad (31)$$

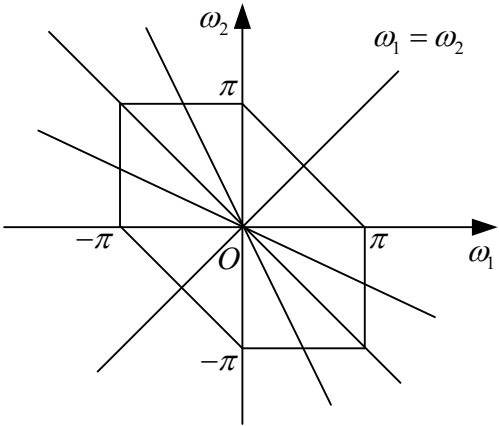

Fig. 6. Symmetric region of bispectrum.

A bispectrum has 12 symmetry regions [24], as shown in Fig. 6. Therefore, it is possible to take a symmetrical area in the result of $B_y(\omega_1,\omega_2)$ for analysis, to completely describe the whole bispectrum. Considering $\{(\omega_1,\omega_2)|0 \leq \omega_1 \leq \omega_2\}$, $B_y(\omega_1,\omega_2)$ is calculated as

$$B_y(\omega_1,\omega_2) = \frac{\pi}{4} i B_1^2 B_2 \delta(\omega_1-2\pi f)\delta(\omega_2-2\pi f) \quad (32)$$

where $\delta(\cdot)$ is the Dirac Delta function, which is described as

$$\begin{cases} \delta(x) = 0 \ (x \neq 0) \\ \int_{-\infty}^{\infty} \delta(x)dx = 1 \end{cases} \quad (33)$$

$\delta(x)$ is a finite maximum at $x = 0$ when the input signal is discretized. So in (32), if and only if $\omega_1 = \omega_2 = 2\pi f$, $B_y(\omega_1,\omega_2)$ is a finite maximum, otherwise it is zero. Therefore, a peak can be observed at the x-y coordinate $(2\pi f, 2\pi f)$ in the 3-dimensional graph of $\{(\omega_1, \omega_2, B_y(\omega_1,\omega_2))|0 \leq \omega_1 \leq \omega_2\}$. Furthermore, considering symmetry, because $(2\pi f, 2\pi f)$ is on the symmetry axis $\omega_1 = \omega_2$, when the area is extended to $\{(\omega_1,\omega_2)|\omega_1 \geq 0, \omega_2 \geq 0\}$ (Area 1 and 2 in Fig. 6), there is also only one peak at $(2\pi f, 2\pi f)$.

The power spectrum of $y(t)$ is derived as

$$P_y(\omega) = \int_{-\infty}^{\infty} f\int_0^{1/f} y(t)y(t+\tau)dt \times e^{-j2\pi\omega}d\tau \quad (34)$$

$$= \frac{1}{2}B_1^2\sqrt{\frac{\pi}{2}}\delta(\omega-2\pi f) + \frac{1}{2}B_2^2\sqrt{\frac{\pi}{2}}\delta(\omega-4\pi f)$$

From (27), (32) and (34), the bicoherence spectrum (in $\{(\omega_1,\omega_2)|\omega_1 \geq 0, \omega_2 \geq 0\}$) can be calculated as

$$bic_y(\omega_1,\omega_2) = \frac{|B_y(\omega_1,\omega_2)|}{\sqrt{P_y(\omega_1+\omega_2)P_y(\omega_1)P_y(\omega_2)}}$$

$$= \frac{\pi}{4}B_1^2 B_2 \delta(\omega_1-2\pi f)\delta(\omega_2-2\pi f) \bigg/ \sqrt{\frac{1}{2}B_1^2\sqrt{\frac{\pi}{2}}\delta(\omega_1-2\pi f) + \frac{1}{2}B_2^2\sqrt{\frac{\pi}{2}}\delta(\omega_1-4\pi f)}$$

$$\bigg/ \sqrt{\frac{1}{2}B_1^2\sqrt{\frac{\pi}{2}}\delta(\omega_2-2\pi f) + \frac{1}{2}B_2^2\sqrt{\frac{\pi}{2}}\delta(\omega_2-4\pi f)}$$

$$\bigg/ \sqrt{\frac{1}{2}B_1^2\sqrt{\frac{\pi}{2}}\delta(\omega_1+\omega_2-2\pi f) + \frac{1}{2}B_2^2\sqrt{\frac{\pi}{2}}\delta(\omega_1+\omega_2-4\pi f)}$$

$$(\omega_1 \geq 0, \omega_2 \geq 0) \quad (35)$$

Considering $\mathbb{F}^{-1}(\delta(\omega)) = \frac{1}{\sqrt{2\pi}}$, it can be proved that if and only if $\omega_1 = \omega_2 = 2\pi f$, the bicoherence spectrum of $y(t)$ reaches its peak, which is calculated as

$$bic_y(2\pi f, 2\pi f) = \frac{\frac{\pi}{4}B_1^2 B_2 \cdot \frac{1}{\sqrt{2\pi}} \cdot \frac{1}{\sqrt{2\pi}}}{\sqrt{\frac{1}{2}B_1^2\sqrt{\frac{\pi}{2}} \cdot \frac{1}{\sqrt{2\pi}} \cdot \frac{1}{2}B_1^2\sqrt{\frac{\pi}{2}} \cdot \frac{1}{\sqrt{2\pi}} \cdot \sqrt{\frac{\pi}{2}} \cdot \frac{1}{2}B_2^2 \cdot \frac{1}{\sqrt{2\pi}} \cdot \sqrt{\frac{\pi}{2}}}} = 1 \quad (36)$$

Obviously, $bic_y(\omega_1,\omega_2) \geq 0$ in (35). Therefore, the range of the corresponding bicoherence value of each x-y coordinate in the bicoherence spectrum is [0,1]. The larger the bicoherence value, the stronger the nonlinear phase coupling between the two frequencies corresponding to the coordinate, that is, the stronger the nonlinearity.

In (30), when $y(t)$ extends to $y(t) = \sum_{n=1}^{+\infty} B_n \sin 2\pi \cdot nft$, similarly, it can be proved that in the 3-dimensional graph of $\{(\omega_1, \omega_2, bic_y(\omega_1,\omega_2))|\omega_1 \geq 0, \omega_2 \geq 0\}$, peaks exist at the x-y coordinate $(i \cdot 2\pi f, j \cdot 2\pi f)(i, j = 1,2,3, ...)$, and their corresponding bicoherence value also equals to 1. Furthermore, it can be proved that when $y(t) = \sum_{n=1}^{+\infty} A_{2n} \cos 2\pi \cdot 2nft + \sum_{n=0}^{+\infty} B_{2n+1}\sin 2\pi \cdot (2n+1)ft$ according to (11), which accurately represents the output of the unilateral saturation hard limit, the conclusion remains the same.

### D. Tricoherence Spectrum and Bilateral Saturation Hard Limit Detection

From the results in II.B, when the self-sustained oscillation occurs, the output $y$ of the bilateral saturation hard limit contains the 3-rd and 5-th harmonics of the oscillation frequency with the same phase as that of the fundamental frequency. Without loss of generality, let its initial phase be 0, i.e.,
$$y(t) = B_1 \sin 2\pi f t + B_3 \sin 2\pi \cdot 3 f t + B_5 \sin 2\pi \cdot 5 f t \quad (37)$$
where $f$ is the oscillation frequency. $B_1$, $B_3$ and $B_5$ are the cosine parts of the magnitude of the fundamental, 3-rd and 5-th Fourier harmonics, respectively.

The Fourier transform is performed three times on the third-order autocorrelation function of $y(t)$ and then the trispectrum of $y(t)$ can be derived as
$$R_f(\tau_1,\tau_2,\tau_3) = f\int_0^{1/f} y(t)y(t+\tau_1)y(t+\tau_2)y(t+\tau_3)dt \quad (38)$$
$$T_y(\omega_1,\omega_2,\omega_3) = \int_{-\infty}^{\infty}\int_{-\infty}^{\infty}\int_{-\infty}^{\infty} R_f(\tau_1,\tau_2,\tau_3) \times e^{-j2\pi(\omega_1+\omega_2+\omega_3)} d\tau_1 d\tau_2 d\tau_3$$

A trispectrum has 96 symmetry regions [24]. Therefore, it is possible to take a symmetrical area in the result of $T_y(\omega_1,\omega_2,\omega_3)$ for analysis, to completely describe the whole trispectrum. Considering $\{(\omega_1,\omega_2,\omega_3)|0 \leq \omega_1 \leq \omega_2 \leq \omega_3\}$, $T_y(\omega_1,\omega_2,\omega_3)$ is calculated as
$$T_y(\omega_1,\omega_2,\omega_3) = \frac{B_1^3 B_3 \pi^{3/2}}{4\sqrt{2}} \delta(\omega_1 - 2\pi f)\delta(\omega_2 - 2\pi f)\delta(\omega_3 - 2\pi f) \quad (39)$$
$$+ \frac{B_1^2 B_3 B_5 \pi^{3/2}}{4\sqrt{2}} \delta(\omega_1 - 2\pi f)\delta(\omega_2 - 2\pi f)\delta(\omega_3 - 6\pi f)$$

If and only if $\omega_1 = \omega_2 = \omega_3 = 2\pi f$ or $\omega_1 = \omega_2 = 2\pi f, \omega_3 = 6\pi f$, $T_y(\omega_1,\omega_2,\omega_3)$ is a finite maximum, otherwise it is zero. Therefore, two peaks can be observed at the x-y-z coordinate $(2\pi f, 2\pi f, 2\pi f)$ and $(2\pi f, 2\pi f, 6\pi f)$ in the 4-dimensional graph of $\{(\omega_1,\omega_2,\omega_3,T_y(\omega_1,\omega_2,\omega_3))|0 \leq \omega_1 \leq \omega_2 \leq \omega_3\}$. To make it easier to generate intuitive graphs, when the area is extended to $\{(\omega_1,\omega_2,\omega_3)|\omega_1 \geq 0, \omega_2 \geq 0, \omega_3 \geq 0\}$, there will be 4 peaks at $(2\pi f, 2\pi f, 2\pi f)$, $(2\pi f, 2\pi f, 6\pi f)$, $(2\pi f, 6\pi f, 2\pi f)$ and $(6\pi f, 2\pi f, 2\pi f)$.

The power spectrum of $y(t)$ is derived as
$$P_y(\omega) = \int_{-\infty}^{\infty} f\int_0^{1/f} y(t)y(t+\tau)dt \times e^{-j2\pi\omega} d\tau$$
$$= \frac{1}{2}B_1^2\sqrt{\frac{\pi}{2}}\delta(\omega - 2\pi f) \quad (40)$$
$$+ \frac{1}{2}B_3^2\sqrt{\frac{\pi}{2}}\delta(\omega - 6\pi f)$$
$$+ \frac{1}{2}B_5^2\sqrt{\frac{\pi}{2}}\delta(\omega - 10\pi f)$$

From (39) and (40), the tricoherence spectrum (in $\{(\omega_1,\omega_2,\omega_3)|\omega_1 \geq 0, \omega_2 \geq 0, \omega_3 \geq 0\}$) can be calculated as
$$tric_y(\omega_1,\omega_2,\omega_3) = \frac{|T_y(\omega_1,\omega_2,\omega_3)|}{\sqrt{P_y(\omega_1+\omega_2+\omega_3)P_y(\omega_1)P_y(\omega_2)P_y(\omega_3)}} \quad (41)$$

In (41), if and only if $\omega_1 = \omega_2 = \omega_3 = 2\pi f$, $\omega_1 = \omega_2 = 2\pi f, \omega_3 = 6\pi f$, $\omega_1 = \omega_3 = 2\pi f, \omega_2 = 6\pi f$ or $\omega_2 = \omega_3 = 2\pi f, \omega_1 = 6\pi f$, the tricoherence spectrum of $y(t)$ reaches its peak.

When $\omega_1 = \omega_2 = \omega_3 = 2\pi f$, it is calculated as
$$tric_y(2\pi f, 2\pi f, 2\pi f) \quad (42)$$
$$= \frac{\frac{B_1^3 B_3 \pi^{3/2}}{4\sqrt{2}} \cdot \frac{1}{\sqrt{2\pi}} \cdot \frac{1}{\sqrt{2\pi}} \cdot \frac{1}{\sqrt{2\pi}}}{\sqrt{\frac{1}{2}B_3^2\sqrt{\frac{\pi}{2}} \cdot \frac{1}{\sqrt{2\pi}} \cdot \frac{1}{2}B_1^2\sqrt{\frac{\pi}{2}} \cdot \frac{1}{\sqrt{2\pi}} \cdot \frac{1}{2}B_1^2\sqrt{\frac{\pi}{2}} \cdot \frac{1}{\sqrt{2\pi}} \cdot \frac{1}{2}B_1^2\sqrt{\frac{\pi}{2}} \cdot \frac{1}{\sqrt{2\pi}} \cdot \sqrt{\frac{\pi}{2}}}}$$
$$= 1$$

When $\omega_1 = \omega_2 = 2\pi f, \omega_3 = 6\pi f$, $\omega_1 = \omega_3 = 2\pi f, \omega_2 = 6\pi f$ or $\omega_2 = \omega_3 = 2\pi f, \omega_1 = 6\pi f$, it is calculated as

$$tric_y(2\pi f, 2\pi f, 6\pi f) \quad (43)$$
$$= \frac{\frac{B_1^2 B_3 B_5 \pi^{3/2}}{4\sqrt{2}} \cdot \frac{1}{\sqrt{2\pi}} \cdot \frac{1}{\sqrt{2\pi}} \cdot \frac{1}{\sqrt{2\pi}}}{\sqrt{\frac{1}{2}B_5^2\sqrt{\frac{\pi}{2}} \cdot \frac{1}{\sqrt{2\pi}} \cdot \frac{1}{2}B_3^2\sqrt{\frac{\pi}{2}} \cdot \frac{1}{\sqrt{2\pi}} \cdot \frac{1}{2}B_1^2\sqrt{\frac{\pi}{2}} \cdot \frac{1}{\sqrt{2\pi}} \cdot \frac{1}{2}B_1^2\sqrt{\frac{\pi}{2}} \cdot \frac{1}{\sqrt{2\pi}} \cdot \sqrt{\frac{\pi}{2}}}}$$
$$= 1$$

Obviously, $tric_y(\omega_1,\omega_2,\omega_3) \geq 0$ in (41). Therefore, the range of the corresponding tricoherence value of each x-y-z coordinate in the tricoherence spectrum is [0,1]. The larger the tricoherence value, the stronger the nonlinear phase coupling among the three frequencies corresponding to the coordinate, that is, the stronger the nonlinearity.

In (37), when $y(t)$ extends to $y(t) = \sum_{n=0}^{+\infty} B_{2n+1} \sin 2\pi \cdot (2n+1)ft$ according to (9), which accurately represents the output of the bilateral saturation hard limit, similarly, it can be proved that in the 4-dimensional graph of $\{(\omega_1,\omega_2,\omega_3,tric_y(\omega_1,\omega_2,\omega_3))|\omega_1 \geq 0, \omega_2 \geq 0, \omega_3 \geq 0\}$, peaks exist at the x-y-z coordinate $((2i+1) \cdot 2\pi f, (2j+1) \cdot 2\pi f, (2k+1) \cdot 2\pi f)(i,j,k = 0,1,2,3,...)$.

*E. Nonlinearity Detection and Classification of VSC Control System*

To sum up, from IV.B, the nonlinearity inside the VSC control system can be detected by transforming the current waveform into $i_d$ and $i_q$ at the terminal of the VSC and performing HOS (i.e., bicoherence/tricoherence) analysis on them. The nonlinearity of $i_d$ represents the nonlinearity in d-axis control loop of the VSC control system, while the nonlinearity of $i_q$ represents the nonlinearity in q-axis.

Combining the conclusions in IV.C and IV.D, Table I is summarized. Essentially, the bicoherence spectrum is applied to detect "phase coupling" in the analyzed signal, which means there exist harmonics whose frequencies $f_1 + f_2 = f_3$ and phases $\varphi_1 + \varphi_2 = \varphi_3$ are satisfied at the same time. Meanwhile, the tricoherence spectrum is applied to detect if $f_1 + f_2 + f_3 = f_4$ and $\varphi_1 + \varphi_2 + \varphi_3 = \varphi_4$ are both satisfied. It is worth noting that the phase equation is actually a sufficient and unnecessary condition of the phase coupling phenomenon, which will be explained in detail in the case study.

As is shown in Table I, the output of the unilateral saturation hard limit contains each order harmonic, so it satisfies both quadratic and cubic phase coupling, which means peaks exist both in its bicoherence and tricoherence spectrums. As a result, the bicoherence spectrum should be first examined, and then the tricoherence spectrum. The nonlinearity can be judged and classified as shown in Fig. 7.

TABLE I
CHARACTERISTICS OF BICOHERENCE AND TRICOHERENCE

|  | Phase Coupling | Unilateral Saturation | Bilateral Saturation |
|---|---|---|---|
| Fourier Series | / | $y(t) = \sum_{n=1}^{+\infty} A_{2n} \cos 2\pi \cdot 2nft + \sum_{n=0}^{+\infty} B_{2n+1} \sin 2\pi \cdot (2n+1)ft$ | $y(t) = \sum_{n=0}^{+\infty} B_{2n+1} \sin 2\pi \cdot (2n+1)ft$ |
| Bicoherence | $f_1 + f_2 = f_3$ $\varphi_1 + \varphi_2 = \varphi_3$ | peaks | no peaks |
| Tricoherence | $f_1 + f_2 + f_3 = f_4$ $\varphi_1 + \varphi_2 + \varphi_3 = \varphi_4$ | peaks | peaks |

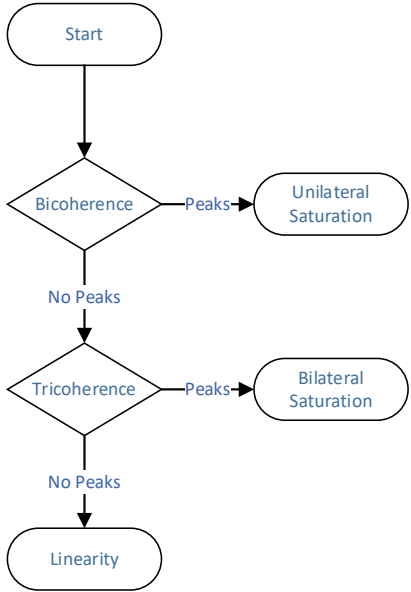

Fig. 7. Flow chart of nonlinearity detection and classification.

## V. CALCULATION PROCESS FOR DNB OF VSC CONTROL SYSTEM

In this section, the specific procedure for applying DNB to the VSC control system in engineering based on the theoretical analysis in the previous section.

To get the time series for calculation, first collect the accident waveform record of current at the terminal of the VSC. The studied signals $x_a(k)$, $x_b(k)$ and $x_c(k)$ are sampled from the three-phase currents $i_a(t)$, $i_b(t)$ and $i_c(t)$. Start sampling from the initial time $t_0$ and set the sampling interval to $\Delta t$. Denote the sampling point with $k$ and the sample length is $L$, i.e.,

$$\begin{cases} x_a(k) = i_a(t_0 + k\Delta t) \\ x_b(k) = i_b(t_0 + k\Delta t) \\ x_c(k) = i_c(t_0 + k\Delta t) \end{cases} \quad k = 1, \cdots, L \quad (44)$$

Step 1: Perform dq transformation on the three-phase sampling signals $x_a(k)$, $x_b(k)$ and $x_c(k)$ and the initial phase $\theta_0$ can be obtained by applying a PLL algorithm.

$$\begin{bmatrix} x_d(k) \\ x_q(k) \\ x_0(k) \end{bmatrix} = \frac{2}{3} \begin{bmatrix} \cos[\theta(k)+\theta_0] & \cos[\theta(k)-\frac{2}{3}\pi] & \cos[\theta(k)+\frac{2}{3}\pi] \\ -\sin[\theta(k)+\theta_0] & -\sin[\theta(k)-\frac{2}{3}\pi] & -\sin[\theta(k)+\frac{2}{3}\pi] \\ \frac{1}{2} & \frac{1}{2} & \frac{1}{2} \end{bmatrix} \begin{bmatrix} x_a(k) \\ x_b(k) \\ x_c(k) \end{bmatrix} \quad (45)$$

where $\theta(k) = \omega_0 k \Delta t + \theta_0$, and $x_d(k)$, $x_q(k)$ and $x_0(k)$ are the transformation results in dq0 coordinate system.

Step 2: Take the transformation result $x_d(k)$ of the previous step as the subsequent signal processing object, i.e.,

$$x(k) = x_d(k) \quad (46)$$

Step 3: Divide $x(k)$ into $M$ segments, and each segment length is $N$ ($L = M \times N$). Record each segment as $x^{(i)}(l)$ ($i = 1, \cdots, M; l = 1, \cdots, N$).

Step 4: Select an appropriate window function, such as a Hanning window, which is described as

$$w(l) = \frac{1}{2}\left[1 + \cos\left(2\pi \cdot \frac{l}{N-1}\right)\right] \quad (47)$$

Multiply each segment of the signal by the window function, and use the obtained results $x'^{(i)}(l)$ for subsequent calculations to reduce leakage errors:

$$x'^{(i)}(l) = x^{(i)}(l) \cdot w(l) \quad (48)$$

Step 5: For each segment $x'^{(i)}(l)$, subtract its mean:

$$\hat{x}^{(i)}(l) = x'^{(i)}(l) - \bar{x}'^{(i)}(l) \quad (49)$$

Step 6: Perform the Fast Fourier Transform (FFT) on each segment $x'^{(i)}(l)$:

$$X_k^{(i)} = \frac{1}{N} \sum_{l=1}^{N} \hat{x}^{(i)}(l) e^{-i2\pi kl/N} \quad k = 1, \cdots, N/2, \ i = 1, \cdots, M \quad (50)$$

Step 7: Deal with the FFT results. Take a small parameter $\sigma$ (such as $\sigma = 0.001$), traverse $i = 1, \cdots, M$, for any $k$, if $X_k^{(i)} < \sigma \max_{k=1,\cdots,N/2}(X_k^{(i)})$, then let $X_k^{(i)} = \sigma^2 \max_{k=1,\cdots,N/2}(X_k^{(i)})$. This step can further increase the difference of the order of magnitude between the white noise and the peak value in the spectrum, so the judgment and analysis of the peak value will not be affected by the appearance of values close to $\frac{0}{0}$ in the area other than the peaks in the bicoherence spectrum.

Step 8: The estimated values of the power spectrum, bispectrum and trispectrum of $x(t)$ are calculated as

$$\hat{P}(m) = \frac{1}{M} \sum_{i=1}^{M} X_m^{(i)} X_m^{*(i)} \quad (51)$$

$$\hat{B}(m,n) = \frac{1}{M} \sum_{i=1}^{M} X_m^{(i)} X_n^{(i)} X_{m+n}^{*(i)} \quad (52)$$

$$\hat{T}(m,n,o) = \frac{1}{M} \sum_{i=1}^{M} X_m^{(i)} X_n^{(i)} X_o^{(i)} X_{m+n+o}^{*(i)} \quad (53)$$

Step 9: Calculate the bicoherence spectrum:

$$\hat{bic}(m,n) = \frac{|\hat{B}(m,n)|}{\sqrt{\hat{P}(m+n)\hat{P}(m)\hat{P}(n)}} \quad (54)$$

Step 10: The obtained bicoherence spectrum is a 3-dimensional graph whose x-y coordinates are the frequencies $(m,n)$, and the $z$ coordinate is the corresponding bicoherence value whose theoretical value range is $[0,1]$.

Step 11: Calculate the tricoherence spectrum:

$$\hat{tric}(m,n,o) = \frac{|\hat{T}(m,n,o)|}{\sqrt{\hat{P}(m+n+o)\hat{P}(m)\hat{P}(n)\hat{P}(o)}} \quad (55)$$

Step 12: Define $\sigma_b$ as the nonlinear threshold (preferably 0.3). A peak in the bicoherence or tricoherence spectrum whose value is greater than $\sigma_b$ (generally close to 1) is considered to characterize the existence of a quadratic or cubic phase coupling, and the coordinates of the peak represent the corresponding frequencies. The judgement and classification of the nonlinearity can be completed following the process in Fig. 7.

Step 13: The steps above implement DNB on the d-axis control loop of the VSC control system. To study the q-axis control loop, back to Step 2, let $x(k) = x_q(k)$ and repeat Step 3-Step 12.

After applying the procedure above, the nonlinear behavior caused by the bilateral or unilateral saturation hard limits in the d- or q-axis control loop can be detected. Among the steps, there are several treatments to increase the resolution and effectiveness of HOS: Step 4 is to reduce spectrum leakage, Step 6 is to eliminate the effect of random phases and Step 7 adds credibility to the presence of peaks.

## VI. CASE STUDY

In this section, three cases will be illustrated and discussed.

Case 1 depicts an artificially constructed signal, which is abstracted from the harmonic characteristics of the unilateral saturation hard limit in II.C, to further discuss if $\varphi_1 + \varphi_2 = \varphi_3$ is a necessary condition of phase coupling in HOS.

Case 2 sets up a grid-connected PMSG model to prove the effectiveness of the proposed process in detecting nonlinearity from a unilateral saturation hard limit by collecting accident waveform records at the terminal of VSCs. In the same case, nonlinearity from a bilateral saturation hard limit is detected using a tricoherence spectrum.

Case 3 sets up an IEEE 9-bus system with three SVGs and one SVC, where the self-sustained oscillation is induced by two SVGs, to demonstrate that HOS can only detect the presence of nonlinearity but can't locate the source of nonlinear oscillation.

### A. Case 1

Considering the following signal:

$$x(n) = \cos\left[2\pi f_1 n / f_s + \varphi_1 + w_1(n)\right]$$
$$+ \cos\left[2\pi f_2 n / f_s + \varphi_2 + w_2(n)\right] \quad (56)$$
$$+ \cos\left[2\pi f_3 n / f_s + \varphi_3 + w_3(n)\right]$$

where $f_1 = 0.6381\ Hz, f_2 = 0.8345\ Hz, f_3 = f_1 + f_2$ and $w_i(n)(i = 1,2,3)$ is -20 dB Gaussian white noise.

In (56), let $\varphi_1 = \varphi_2 = \varphi_3 = 0$, recorded as $x_1(n)$. Then, let $\varphi_1 = \varphi_2 = 0, \varphi_3 = \pi/2$, recorded as $x_2(n)$.

Fig. 8(a) is the frequency spectrum of $x_1(n)$, from which three frequency components $f_1, f_2$ and $f_3$ can be found, but the relationship among them can't be determined. Fig. 8(b) is the bicoherence spectrum of $x_1(n)$. Its peak appears at $(f_1, f_2)$ and $(f_2, f_1)$, and the peak value is 1.0, which means the power of the frequency component $f_3$ entirely comes from quadratic phase coupling of $f_1$ and $f_2$, which proves the conclusion in IV.C.

Fig. 8(c) and Fig. 8(d) are the frequency spectrum and the bicoherence spectrum of $x_2(n)$, which are almost the same as Fig. 8(a) and Fig. 8(b), respectively. In our initial setting, however, $\varphi_1 + \varphi_2 = \varphi_3$ in $x_1(n)$, while $\varphi_1 + \varphi_2 \neq \varphi_3$ in $x_2(n)$.

Most of the previous researches [25]–[27] consider quadratic phase coupling equivalent to $f_1 + f_2 = f_3$ and $\varphi_1 + \varphi_2 = \varphi_3$, which is actually a sufficient and unnecessary condition. As long as $\varphi_1 + \varphi_2$ keeps a fixed difference with $\varphi_3$, i.e., $\varphi_1 + \varphi_2 - \varphi_3$ is constant, the phase coupling exists and peaks can be seen in the bicoherence spectrum. This conclusion is important because according to (11), the output of the unilateral saturation hard limit can be expressed as

$$y(t) = \sum_{n=1}^{+\infty} A_{2n} \cos 2\pi \cdot 2nft + \sum_{n=0}^{+\infty} B_{2n+1} \sin 2\pi \cdot (2n+1) ft \quad (57)$$
$$= \sum_{n=1}^{+\infty} A_{2n} \sin\left(2\pi \cdot 2nft + \frac{\pi}{2}\right) + \sum_{n=0}^{+\infty} B_{2n+1} \sin 2\pi \cdot (2n+1) ft$$

Therefore, its peaks appear in the bicoherence spectrum like a "chessboard", because any two integer multiples of the fundamental frequency have the property of quadratic phase coupling. Otherwise, if the bicoherence can only detect those satisfying $f_1 + f_2 = f_3$ and $\varphi_1 + \varphi_2 = \varphi_3$ at the same time, there will be no peaks in the bicoherence spectrum at all, which does not match the actual situation.

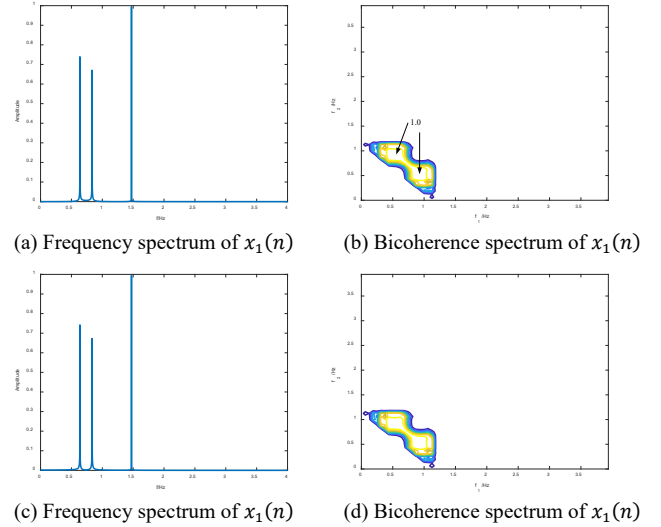

(a) Frequency spectrum of $x_1(n)$  (b) Bicoherence spectrum of $x_1(n)$

(c) Frequency spectrum of $x_1(n)$  (d) Bicoherence spectrum of $x_1(n)$

Fig. 8. Detection of quadratic phase coupling.

### B. Case 2

Set up a detailed grid-connected PMSG model in PSCAD/EMTDC, the structure of the VSC control system is shown in Fig. 9. Adjust the parameters to make the hard limit of the PI in the d-axis outer control loop of voltage take effect. The setting of the parameters is listed in Table II.

The simulation is implemented as follows:

$t = 0.0\ s$: use a voltage source to charge the DC capacitor; in the initial state, the PMSG is off-grid and the active power and reactive power references are both 0.

$t = 0.2\ s$: the DC capacitor side is switched to the power source, and the PMSG is connected to the grid.

$t = 1.0\ s$: the active power is set to 0.34 MW.

$t = 2.0\ s$: $G_i(s)$ is set to $0.2 + 20/s$.

$t = 4.0\ s$: $G_i(s)$ is set to $0.012 + 12.5/s$.

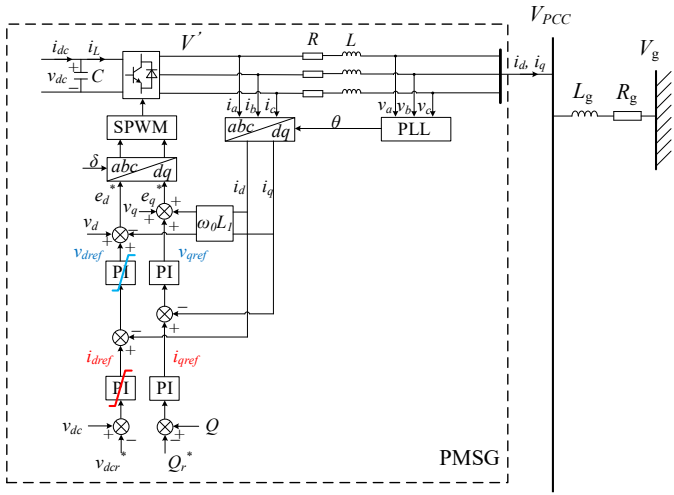

Fig. 9. The PMSG model for DNB.

TABLE II
PARAMETERS IN CONTROL BLOCKS OF PMSG

| Symbol | Description | Value |
| --- | --- | --- |
| $V_g$ | Grid line voltage | 0.69 kV |
| $f_0$ | Fundamental frequency | 50 Hz |
| $P_N$ | Rated capacity of PMSG | 1.5 MW |
| $H_{PLL}(s)$ | Phase-locked loop | $500 + 900/s$ |
| $P$ | Active power | 0.34 MW |
| $C$ | DC link capacitor | 200 mF |
| $R$ | Connection resistance | 0.001 Ω |
| $L$ | Connection inductance | 0.35 mH |
| $R_g$ | Grid-side resistance | 0.005 Ω |
| $L_g$ | Grid-side inductance | 0.4 mH |
| $G_{dc}(s)$ | DC-voltage controller | $9 + 500/s$ |
| $G_q(s)$ | Reactive power controller | $0.3 + 50.28/s$ |
| $G_i(s)$ | Inner-loop current controller | $0.012 + 12.5/s$ |

Fig. 10 is the current reference $i_{dref}$. With the parameters listed in Table II, after $t = 4.0\ s$, the system has a pair of characteristic roots on the right side of the imaginary axis, which induces a divergent oscillation. When the oscillation causes $i_{dref}$ to reach the hard limit of the PI in the d-axis outer control loop of voltage, it becomes unilateral saturated. Meanwhile, at PCC, an equal-amplitude self-sustained oscillation of 33.8Hz can be observed, as is shown in Fig. 11. Considering the transient process is short, the collected accident waveform record may only contain the equal-amplitude part, which is of superficial resemblance with the linear weakly-damped oscillation.

Fig. 12 is the bicoherence and power spectrums of $i_{dref}$ and $i_{dpcc}$. The bicoherence spectrums are figured as contour maps, whose x-y axis range is restricted in $\{(f_1, f_2)|\omega_1 \geq 0, \omega_2 \geq 0\}$. As is shown in Fig. 12(a), the bicoherence spectrum of $i_{dref}$ presents as a "chessboard", which means quadratic phase coupling exists between any two integer multiples of the fundamental frequency. Normally, the transfer function of the VSC control system is lowpass, which can be seen by comparing Fig. 12(b) and Fig. 12(d). The bicoherence spectrum of $i_{dpcc}$, however, still keeps the property of quadratic phase coupling, as is shown in Fig. 12(c). It is acceptable that the peaks of higher harmonics disappear because their amplitudes are too small to maintain distinction from background noise. Therefore, Fig. 12(c) proves that nonlinearity from a unilateral saturation hard limit can be detected by collecting accident waveform records at the terminal of VSCs based on HOS analysis. Especially in this case, nonlinearity exists in the d-axis control loop of the VSC control system.

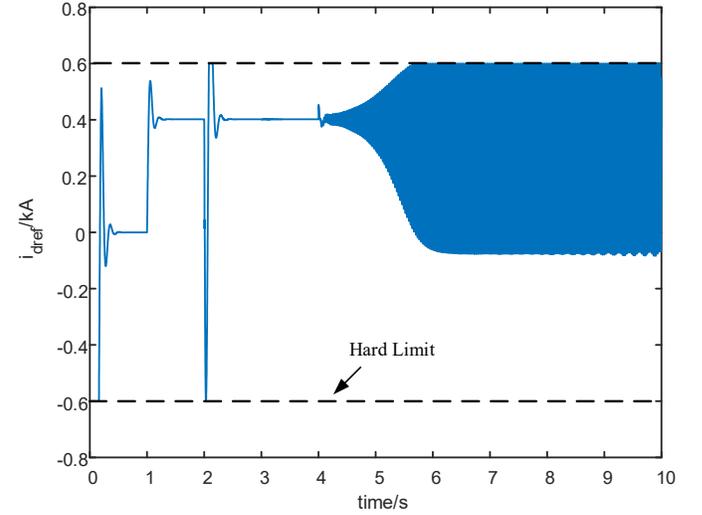

Fig. 10. Current reference in $d$ axis.

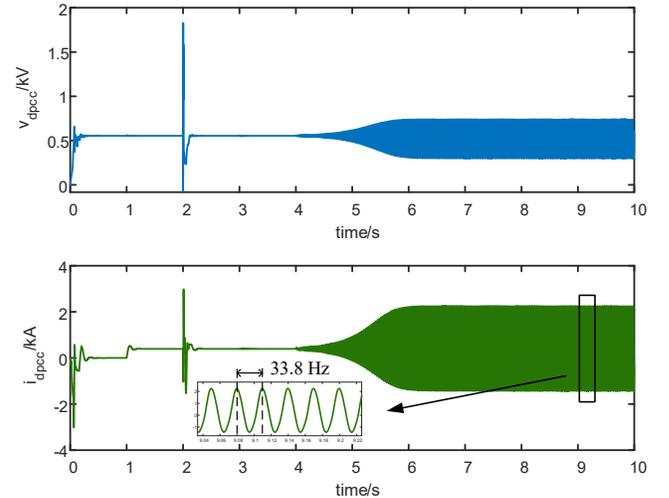

Fig. 11. Voltage and current at PCC in $d$ axis.

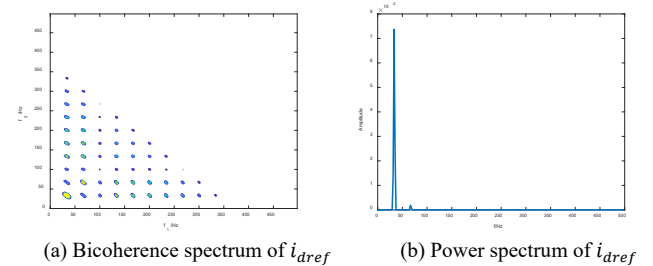

(a) Bicoherence spectrum of $i_{dref}$  (b) Power spectrum of $i_{dref}$

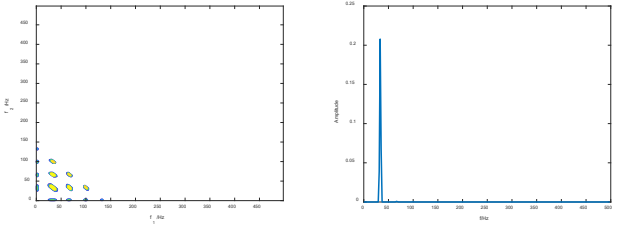

(c) Bicoherence spectrum of $i_{dpcc}$  (d) Power spectrum of $i_{dpcc}$

Fig. 12.  Bicoherence and power spectrum of $i_{dref}$ and $i_{dpcc}$.

In the same case, unloose the hard limit of the PI in the d-axis outer control loop of voltage, and set the upper and lower limit of the PI in the d-axis inner control loop of current $v_{dref}$ to $\pm 0.08$, which takes effect and induces a bilateral saturation sustained oscillation.

Fig. 13(a) shows the bicoherence spectrum of $i_{dpcc}$, which has no peaks in the contour maps, with the global maximum value only $0.002134 (\ll 1)$, meaning there's no unilateral saturation. Fig. 13(b) shows the tricoherence spectrum of $i_{dpcc}$, which has a peak at $(33.8Hz, 33.8Hz, 33.8Hz)$. The aliasing expands the peak range and makes the maximum value larger than 1, but it still clearly points out the bilateral saturation in the system, which proves the effectiveness of the calculation and classification process proposed in IV.E and V.

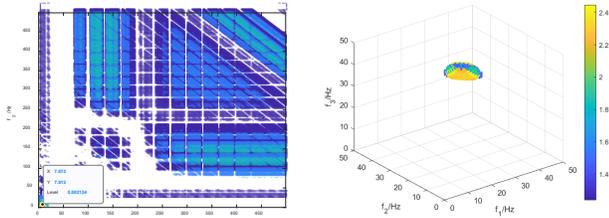

(a) Bicoherence spectrum of $i_{dpcc}$  (b) Tricoherence spectrum of $i_{dpcc}$

Fig. 13.  Bicoherence and tricoherence spectrum of $i_{dpcc}$.

### C. Case 3

Fig. 14 illustrates the topology for the case-study system deriving from the IEEE 9-bus system, where the parameters of the network composed of buses 1 to 9 are consistent with the IEEE 9-bus system. Different from the IEEE benchmark system, an SVC is located at medium-voltage bus 13 for reactive power adjustment, and two grid-connected wind farms and their corresponding SVGs are located at the bus 14 and bus 15, respectively. The PMSGs and SVGs adopt the double-loop VSC control as shown in IV.A, and the parameters of control are listed in Table III.

The case-study system is implemented by PSCAD/EMTDC simulation, and the self-sustained oscillation is induced by the mismatch of the reference terminal voltage of $SVG_1$ and $SVG_2$, which are denoted by $V_{ref1}$ and $V_{ref2}$, respectively. Fig. 15 shows the oscillation-related waveforms of voltages and currents. In Fig. 15(a), when $t < 2 \ s$, set $V_{ref1} = V_{ref2}$, and the voltage amplitudes $V_{14}$, $V_{PCC1}$, and $V_{PCC2}$ are stable; by contrast, when $t \geq 2.45 \ s$, set $V_{ref1} \neq V_{ref2}$, then the voltage amplitude $V_{14}$ deviates from the previous equilibrium point, which fluctuates and ranges from 0.97 to 1.04 pu. Meanwhile, according to Fig. 15 (b), the current amplitude of $i_{14}$ increases significantly when $t \geq 2.45 \ s$. The major harmonic frequencies of instantaneous current $i_{14}$ are $17.5 \ Hz$ and $82.5 \ Hz$ ($50 \pm 32.5 \ Hz$), so that it is demonstrated that there is a sub-synchronous current injection flowing from VSCs to networks. As is shown in Fig. 15(c), the static VAR compensator (SVC) is stimulating a linear LC resonance, whose combination of the capacitive reactance and the network inductance matches the sub-synchronous frequency of the VSCs. The figure shows the waveform of current for SVC, and it is demonstrated that the SVC works as a harmonic amplifier.

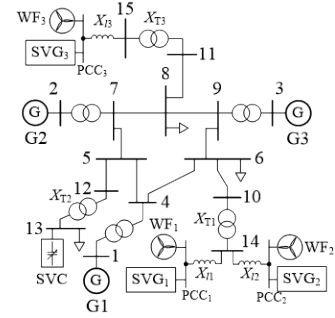

Fig. 14.  The topology of the case study system.

TABLE III
PARAMETERS OF NETWORK AND VSC CONTROL FOR CASE-STUDY SYSTEM

| Parameters | Values |
|---|---|
| SVG voltage control loop ($K_{Pvd}$, $K_{Ivd}$, $K_{Pvq}$, $K_{Ivq}$) | 2.5 pu, 1000 pu, 2 pu, 20 pu |
| Reference of terminal voltage control ($V_{ref1}$, $V_{ref2}$, $V_{ref3}$) | 1.005 pu, 1.005 pu, 1.005 pu ($t < 2$ s) 1.005 pu, 1.000 pu, 1.005 pu ($t \geq 2$ s) |
| Current control loop ($K_{Pi}$, $K_{Ii}$) | 40 pu, 6250 pu |
| Connection impedance ($X_{l1}$, $X_{l2}$, $X_{l3}$) | 0.0051 pu, 0.0038 pu, 0.0256 pu |
| Line resistance ($R_{6-10}$, $R_{8-11}$) | 0.0017 pu, 0.0054 pu |
| Line impedance ($X_{6-10}$, $X_{8-11}$) | 0.0092 pu, 0.0178 pu |
| Transformer impedance ($X_{T1}$, $X_{T2}$, $X_{T3}$) | 0.0586 pu, 0.0586 pu, 0.0576 pu |

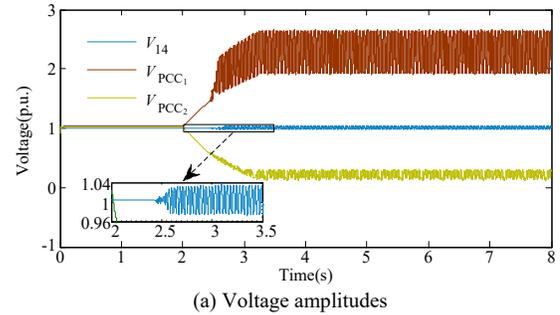

(a) Voltage amplitudes

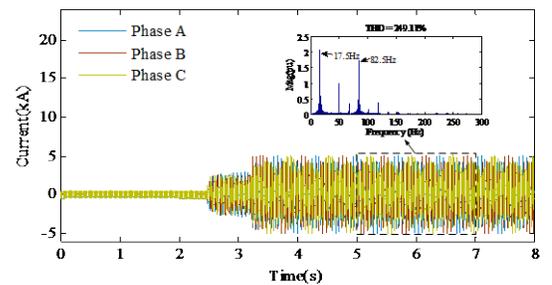

(b) Three-phase currents $i_{14}$ and the corresponding spectrum.

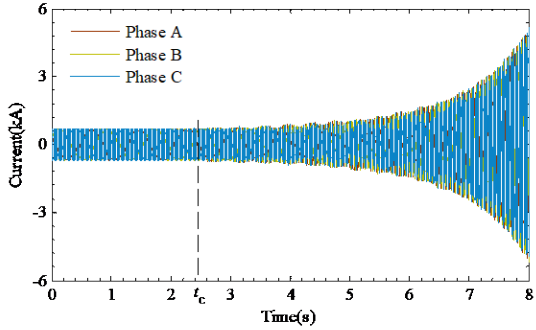

(c) Three-phase currents of SVC

Fig. 15. Waveforms of voltages and currents.

TABLE IV
VALUE COMPARISON OF $\mu$ FOR HARMONIC SOURCE SEARCHING

| Index | $SVG_1$ | $SVG_2$ | $SVG_3$ | SVC |
|---|---|---|---|---|
| $\mu$ (p.u.) | 0.3727 | 0.1598 | 0.1043 | 0.0079 |

A nonlinear index $\mu$ for checking the flatness of bicoherence spectrum could be defined as [28]

$$\mu \triangleq \left| \hat{bic}^2_{max} - \left( \overline{\hat{bic}^2} + 2\sigma_{\hat{bic}^2} \right) \right| \quad (57)$$

where, $\overline{\hat{bic}^2}$ is the average of the estimated squared bicoherence; $\sigma_{\hat{bic}^2}$ is the standard deviation of $\hat{bic}^2$. In (57), if $\mu \gg 0$, the signal generating process is nonlinear.

Table IV reports that the $\mu$ values for SVGs are significantly higher than that of SVC, meaning that the nonlinearity is detected in the system and the SVGs are the major contributor to the nonlinearity. However, hard limits take effects in only two of the SVGs. So, the proposed method can only detect the existence of nonlinear behavior in the system, but it cannot locate it. Precise localization requires the introduction of other methods, which will be elaborated in another article.

## VII. CONCLUSION AND DISCUSSION

This paper proposes a method based on HOS analysis for DNB of the VSC control system in wind farms, where PMSGs and VSGs are modeled as a unified VSC control model. The paper establishes a corresponding relationship between the bicoherence spectrum and the unilateral saturation hard limit, as well as the tricoherence spectrum and the bilateral saturation hard limit. Based on it, the DNB and classification of the VSC control system are studied and proved, and the detailed calculation and estimation process is proposed. The bicoherence spectrum and the tricoherence spectrum actually look for the quadratic and cubic phase coupling in the analyzed signal, which is further illustrated in the case study part.

Further work may include two aspects. Firstly, the HOS analysis can be applied to no matter what the measurement signal is, but whether the proposed method can be extended to any equipment needs further examined. This paper proves the applicability in the VSC, which exists in energy storage equipment and DC transmission systems in addition to wind power systems. Other devices may have other parts from the hard limit to the terminal, and whether the effects of these parts can be eliminated is not demonstrated in this paper. Secondly, as is discussed in VI.C, DNB is not enough to form an effective control measure and generator tripping strategy when a self-sustained oscillation accident occurs. So, a nonlinear oscillatory source localization method needs to be further studied.

## APPENDIX

TABLE A
N-TH COMPONENTS OF BILATERAL SATURATION HARD LIMIT OUTPUT

| $n$ | $A_n$ | $B_n$ |
|---|---|---|
| 0 | 0 | 0 |
| 1 | 0 | $\dfrac{2a\sqrt{1-\frac{a^2}{A^2}} + 2A\sin^{-1}\frac{a}{A}}{\pi}$ |
| 2 | 0 | 0 |
| 3 | 0 | $\dfrac{4a(1-\frac{a^2}{A^2})^{3/2}}{3\pi}$ |
| 4 | 0 | 0 |
| 5 | 0 | $\dfrac{4a\sqrt{1-\frac{a^2}{A^2}}(8a^4 - 11a^2A^2 + 3A^4)}{15A^4\pi}$ |
| 6 | 0 | 0 |
| 7 | 0 | $\dfrac{48a\cos\left(7\sin^{-1}\frac{a}{A}\right) + 28A\sin\left(6\sin^{-1}\frac{a}{A}\right) - 21A\sin\left(8\sin^{-1}\frac{a}{A}\right)}{84\pi}$ |

TABLE B
N-TH COMPONENTS OF UNILATERAL SATURATION HARD LIMIT OUTPUT

| $n$ | $A_n$ | $B_n$ |
|---|---|---|
| 0 | $a + 2A_0 - \dfrac{2}{\pi}\sqrt{1-\dfrac{a^2}{A^2}}A - \dfrac{2a}{\pi}\sin^{-1}\dfrac{a}{A}$ | |
| 1 | 0 | $\dfrac{2a\sqrt{1-\frac{a^2}{A^2}} + A\pi + 2A\sin^{-1}\frac{a}{A}}{2\pi}$ |
| 2 | $\dfrac{2\sqrt{1-\frac{a^2}{A^2}}(-a^2 + A^2)}{3A\pi}$ | 0 |
| 3 | 0 | $\dfrac{2a(1-\frac{a^2}{A^2})^{3/2}}{3\pi}$ |
| 4 | $\dfrac{2\sqrt{1-\frac{a^2}{A^2}}(6a^4 - 7a^2A^2 + A^4)}{15A^3\pi}$ | 0 |
| 5 | 0 | $\dfrac{2a\left(3 + \frac{8a^4}{A^4} - \frac{11a^2}{A^2}\right)\sqrt{1-\frac{a^2}{A^2}}}{15\pi}$ |
| 6 | $\dfrac{2\sqrt{1-\frac{a^2}{A^2}}}{105A^5\pi}\left(-80a^6 + 128a^4A^2 - 51a^2A^4 + 3A^6\right)$ | 0 |
| 7 | 0 | $\dfrac{1}{168\pi}\left(48a\cos 7\sin^{-1}\dfrac{a}{A} + 28A\sin 6\sin^{-1}\dfrac{a}{A} - 21A\sin 8\sin^{-1}\dfrac{a}{A}\right)$ |

**Zetian Zheng** (S'18) received the B.Eng. degree in Electrical Engineering from Tsinghua University, Beijing, China, in 2017. He has been a Ph.D. student in the Department of Electrical Engineering, Tsinghua University since 2017. His area of research is analysis and control of the power system with renewable energy sources.

**Chen Shen** (M'98-SM'07) received the B.E. and Ph.D. degrees in electrical engineering from Tsinghua University, Beijing, China, in 1993 and 1998, respectively. From 1998 to 2001, he was a Postdoc Research Fellow with the Department of Electrical Engineering and Computer Science, University of Missouri Rolla, Rolla, MO, USA. From 2001 to 2002, he was a Senior Application Developer with ISO New England, Inc., MA, USA. Since 2009, he has been a Professor with the Department of Electrical Engineering, Tsinghua University. He is currently the Associate Director of the State Key Laboratory of Power System and Generation Equipment and the Director of the Center of Cloud-Based Simulation and Intelligent Decision-Making (CSAID), Sichuan Energy Internet Research Institute, Tsinghua University. His research interests include power system analysis and control, renewable energy generation, and smart grids. He is the author/coauthor of more than 180 technical papers and one book, and holds 32 issued patents.